\documentstyle[twocolumn,aps,epsfig,here]{revtex}

\begin{document}

\title{Charged particle multiplicity near mid-rapidity in central Au+Au collisions \\
at $\sqrt{s} =$ 56 and 130 AGeV }
\author{ B.B.Back$^1$, M.D.Baker$^2$,
D.S.Barton$^2$, S.Basilev$^5$, R.Baum$^8$,R.R.Betts$^{1,7}$, A.Bia\l as$^4$, 
R.Bindel$^8$, W.Bogucki$^3$, 
A.Budzanowski$^3$
% ...
\\
(PHOBOS collaboration)\\
$^1$~Physics Division, Argonne National Laboratory, Argonne, IL 60439-4843\\
$^2$~Chemistry and C-A Departments, Brookhaven National Laboratory, Upton, NY 11973-5000\\
$^3$~Institute of Nuclear Physics, Krak\'{o}w, Poland\\
$^4$~Department of Physics, Jagellonian University, Krak\'{o}w, Poland\\
$^5$~Laboratory for Nuclear Science, Massachusetts Institute of Technology, Cambridge, MA  02139-4307\\
$^6$~Department of Physics, National Central University, Chung-Li, Taiwan\\
$^7$~Department of Physics, University of Illinois at Chicago, Chicago, IL 60607-7059\\
$^8$~Department of Chemistry, University of Maryland, College Park, MD 20742\\
$^9$~Department of Physics and Astronomy, University of Rochester, Rochester, NY 14627\\
$^{10}$~Department of Physics, Yale University, New Haven, CT 06520
}

\date{\today}
\maketitle

\begin{abstract}\noindent
We present the first measurement of pseudorapidity densities of 
primary charged particles near mid-rapidity in Au+Au collisions 
at $\sqrt{s} =$ 56 and 130 AGeV. For the most central collisions,
we find the charged particle pseudorapidity density to be 
$dN/d\eta |_{|\eta|<1} = 408 \pm 12 \mbox{(stat)} \pm 30 \mbox{(syst)}$ 
at 56 AGeV and $555 \pm 12 \mbox{(stat)} \pm 35 \mbox{(syst)}$ at 130 AGeV,
values that are higher than any previously observed in nuclear collisions. 
Compared to proton-antiproton collisions, our data show an increase in 
the pseudorapidity density per participant by more than 40\%
at the higher energy.
\end{abstract}


In June 2000, the Relativistic Heavy-Ion Collider (RHIC) at Brookhaven National 
Laboratory delivered the first collisions
between Au nuclei at the highest center of mass energies achieved 
in the laboratory to date.  
In this paper we present data taken with the PHOBOS detector 
during the first collider run at energies of $\sqrt{s}$ = 56 
and 130 AGeV.  
The ultimate goal of our work is to understand the behavior of 
strongly interacting matter at conditions of extreme density and 
temperature. Quantum chromodynamics (QCD),
the fundamental theory of strong interactions, predicts that for sufficiently 
high energy density a new state of matter will be formed, the so-called
quark-gluon plasma (QGP) \cite{qgp}. The measurements shown here represent the 
first step toward the development of a full picture of the dynamical
evolution of nucleus-nucleus collisions at RHIC energies.

Studying the dependence of charged particle densities 
on energy and system size provides information on the interplay
between hard parton-parton scattering processes, which can be calculated 
using perturbative QCD, 
and soft processes, which are treated by phenomenological models 
that describe the non-perturbative sector of QCD.
Predictions for multi-particle production in high-energy heavy-ion 
collisions, obtained  from a variety of models, 
typically vary by up to a factor of two \cite{qm99}. 

In this letter we report data for the most central Au+Au collisions
detected in our apparatus. We have determined the energy dependence of
the density of primary charged particles emitted 
near $90^\circ$ to the beam axis, characterized by 
the pseudorapidity density $dN/d\eta |_{|\eta|<1}$, 
where $\eta = - \ln \tan(\theta/2)$ and  $\theta$ is the polar 
angle from the beam axis. 
These data provide the 
first means to constrain models of heavy-ion collisions at RHIC energies.
They will allow the extraction of  basic information about the 
initial conditions in these collisions, in particular the 
energy density, and thus form an essential element for the
proper prediction or description of other observables. 

\begin{figure}[t]
\centerline{
\epsfig{file=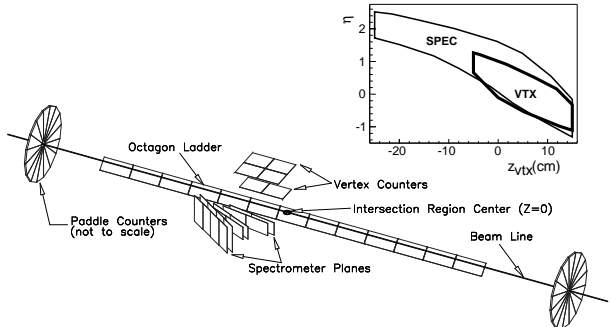,width=8cm}}
\caption{Detector setup for the initial running period. The insert shows 
the $\eta$ acceptance of the SPEC and VTX subdetectors as a function of primary 
vertex position.}
\label{1}
\end{figure}

The PHOBOS detector employs silicon pad detectors to perform tracking, 
vertex detection and multiplicity measurements.
Details of the setup and the 
layout of the silicon sensors can be found elsewhere\cite{phobos1,phobos2}. 
For the initial running
period of the accelerator only a small fraction of the full setup was 
installed.  It included the first 6 layers of the silicon
spectrometer (SPEC), part of the two-layer silicon vertex detector (VTX) and
one ladder of the large acceptance Octagon multiplicity detector (see Fig.~\ref{1}).
In total, the installed sensors had
20000 readout channels, of which less than 2\% were non-functional.
The detector setup also included two sets of 16 scintillator 
counters (``paddle counters'') located at distances of $-3.21$~m (PN) and 
$3.21$~m (PP) from the nominal interaction  point along the 
beam ($z$) axis. These counters subtended pseudorapidities between
$3 < |\eta |< 4.5$. They served as the primary event trigger and
were used for event selection.
Two zero-degree calorimeters (ZDCP,ZDCN) at $z = \pm 18.5$~m 
provided additional information for event selection by measuring the energy
deposited by spectator neutrons. 

Monte Carlo (MC) simulations of the detector performance 
were based on the HIJING event generator \cite{hijing} 
and the GEANT~3.21 simulation package, folding in
the signal response for scintillator counters and silicon sensors.

In this analysis, we used the data from the SPEC subdetector 
to reconstruct tracks 
and to obtain the position, $z_{vtx}$, of the collision vertex. 
The SPEC and VTX subdetectors
were then used to count the multiplicity of charged particles near
$\eta = 0$. Their acceptance, covering $-1 < \eta < 2.5$, is 
shown in the insert in Fig.~\ref{1}. The 
strong dependence on $z_{vtx}$ allows
us  to check our understanding of acceptance and backgrounds
by testing the stability of our results as a function of $z_{vtx}$.

The readout trigger selected events based on the coincidence of at
least one hit in both PP and PN within a time window of 38~ns.  This condition selected
most collision events as well as various backgrounds.
Offline event selection was achieved by requiring the paddle
time difference to be less than 8~ns,
corresponding to a maximum displacement of roughly $\pm 120$~cm 
relative to $z = 0$. 
Additional rejection of residual background can be achieved by
requiring the ZDC time difference to be less than 20~ns.  
Since the ZDCs only detect spectator neutrons, they are
slightly inefficient for the most central events.  
Thus, we accepted events with good ZDC timing or high multiplicity in the paddles
or both.
%
Double beam backgrounds were studied by
using experimental runs where the beams did not collide.
We found that the overall rate was less than 1\% of all events and that 
there was no background mis-identified as central events,
which are characterized by large signals in both paddle counters.

To select the most central events we have used an estimator based on the mean of
gain-normalized ADC values in the 16 scintillator counters in PP and PN.  
Primary charged particles each leave approximately 1.8
MeV in the scintillator.  
Slow secondary particles that traverse the counters at large 
angles may deposit larger amounts of energy, and thus mimic 
a larger multiplicity.
To reduce this effect, for each event we discarded the four scintillators
in each set of paddle counters with the largest signals.  The average of
the pulse heights of the remaining 12 scintillators, PP$_{12}$ and PN$_{12}$, 
was then calculated for each set.  Finally, we used the paddle mean 
$\frac{1}{2}$ (PP$_{12}$ +PN$_{12}$) as an observable proportional to
the number of particles hitting the paddle counters.  

\begin{figure}[t]
\centerline{
\epsfig{file=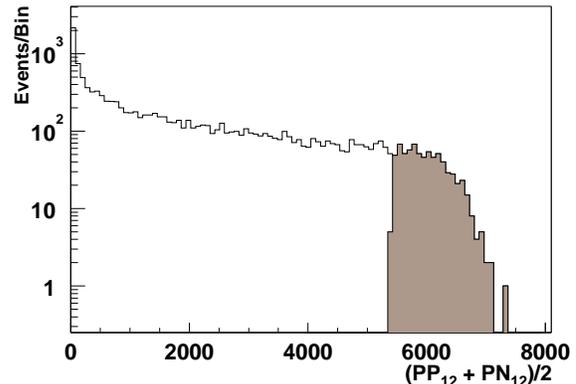,width=8cm}
}
\caption{Paddle signal distribution for data events at $\sqrt{s} = $ 130~AGeV.}
\label{3}
\end{figure}

Assuming that the number of particles 
hitting the paddle counters increases monotonically with  
increasing number of participants,
a cut on large paddle mean selects the events with the largest
number of participants, in both the data and simulation.  
This has been confirmed with the ZDCs for our more central events.
We observe a good anti-correlation between the paddle mean 
and the total energy deposited in the ZDCs.
For this analysis central events were selected
by choosing the 6\% of events with the largest paddle mean (see Fig.~\ref{3}). 
We obtained a sample of 382 events for $\sqrt{s} = 56$~AGeV and 
724 events for $\sqrt{s} = 130$~AGeV.
We estimated the average number of participants
$\langle N_{part} \rangle$ chosen by the 6\% cut using 
simulations \cite{hijing}.
MC studies showed that up to 10\% of the inelastic
cross section fails to produce enough particles to satisfy our trigger
conditions.
We also  studied the distribution of $PP_{12}$
after a 6\% cut on $PN_{12}$ and vice-versa.  This showed 
the effect of additional fluctuations introduced by 
slow secondary particles.
Considering both of these effects, we estimate that the
systematic uncertainty in $\langle N_{part} \rangle$ is 5\%.
Applying the 6\% centrality cut to MC events we deduce that the number of 
participants in the data is $\langle N_{part} \rangle = 330 \pm 4$
(stat) 
for $\sqrt{s} = $56~AGeV and $343 \pm 4$ (stat) for 130~AGeV. 
The RMS of the distribution of $N_{part}$ for the MC events is 27
at 56 AGeV and 25 at 130 AGeV.

The analysis proceeded as follows: Using the 
information from the SPEC sensor planes a road-following tracking
algorithm reconstructed straight line tracks passing through at least 4 layers
of the detector and having a $\chi^2$-probability of better than 1\% for 
a non-vertex constrained straight line fit. MC studies show that in this
step the tracking efficiency is better than 90\%, with less than
10\% ghost tracks. For central events, on average 13 tracks were
reconstructed at $\sqrt{s} = 56$~AGeV and 18 tracks at 130~AGeV.

The position of the primary collision vertex was determined as the point 
of closest approach for the found tracks. 
The distribution of collision vertices, which define the beam orbit, was 
found to be very stable within each data set, while an offset of 0.7~mm
between the two beam energies was observed. To optimize the stability 
of the vertex finding algorithm, which is essential for the multiplicity
determination, we performed a second track finding analysis of the data by
confining the vertex position to within 3~mm distance to the previously determined
beam orbit in the transverse direction and within  
$-25 < z_{vtx} < 15$~cm longitudinally. 
For central events in this region, MC simulations show that the vertex
finding algorithm is more than 99\% efficient. As a cross-check of 
the vertex finding procedure we compared the 
pointing accuracy of the found tracks to the fitted vertex for data and MC 
and found a most probable  value of  1~mm in both cases.

Within the vertex fiducial volume we found a total of 103
central events for the low energy and 151 central events in the high energy data set.
For the selected events, the multiplicity of charged particles was determined
 using the SPEC and VTX subdetectors independently. The measurement was 
done by counting tracklets, which are two hit combinations in 
consecutive layers of the SPEC or  VTX subdetector consistent with 
a track originating at the primary vertex. For every hit the coordinates
in $\eta$ and $\phi$ were calculated relative to the fitted vertex 
position, where $\phi$ is the azimuthal angle in radians.
Then we determined for each hit in the first layer the closest hit 
in the second layer.
Finally, hit combinations with a distance of $\sqrt{\delta \eta ^2 + 
\delta \phi^2 } < 0.015$ in the SPEC subdetector  and 
$|\delta \eta| < 0.1$ in the VTX subdetector were retained as tracklets.
To obtain a constraint on possible random channel-by-channel 
inefficiencies in the sensors and readout-chain, we compared
the multiplicities of single hits in VTX and SPEC, two-hit tracklets 
in VTX and SPEC and tracks in SPEC.
A detection efficiency of less than 100\% would enter in different
powers into the different measurements.
We find no evidence for inefficiencies beyond the $<2$\% of dead channels 
identified in our standard calibration procedure.

MC studies show that the number of tracklets per event near $\eta=0$ is 
approximately proportional to the number of primary charged particles 
per event. Here we define as primary particles all charged hadrons produced
in the collision, including the  products of strong and electromagnetic
decays, but excluding feed-down products from weak decays and hadrons 
produced in secondary interactions.

The pseudorapidity density of primary charged
particles, $dN/d\eta |_{|\eta|<1}$, was obtained from the observed 
distribution of tracklets, $dN/d\eta (\mbox{tracklets})$,
by accounting for several factors: The efficiency of 
tracklet finding, background due to noise hits, 
secondary and feed-down particles and the 
acceptance.
The correction was done using the following procedure: 
Events from a standard event generator were propagated 
through the GEANT simulation of our setup.
The simulated data were then processed with the identical
reconstruction chain as the real data, including the event selection
based on the paddle counter signals.
For each simulated event we determined the number of tracklets, $N_{tracklets}$, found
in the region $|\eta| < 1$ and the number of primary charged particles $N_{primaries}$ 
in the same pseudorapidity region. 
We then determined the vertex dependent proportionality factor $\alpha(z_{vtx})$ 
by calculating  $\langle N_{tracklets}/N_{primaries}\rangle$ as a function of $z_{vtx}$.
\begin{figure}[t]
\centerline{ \epsfig{file=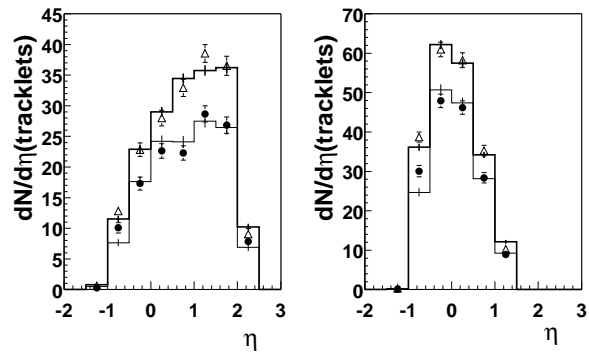,width=8cm} }
\caption{Tracklet pseudorapidity density in the detector acceptance per event for 
data at $\sqrt{s} = $ 56 (circles) and 130~AGeV (triangles) for 
SPEC (left) and VTX (right),
in comparison to scaled HIJING simulations (solid lines).}
\label{5}
\end{figure}

We repeated this procedure using both the HIJING and 
VENUS \cite{venus} event generators, which 
predict multiplicities that differ by a factor 2-3. We also compared $\alpha(z_{vtx})$
obtained for running the event generators at various different energies.
For all MC runs, $\alpha$ agreed to better than 5\%, demonstrating that 
the tracklet counting procedure provides a robust measure of the primary 
charged particle pseudorapidity density.

MC studies showed that less than 5\% of the particles 
emitted into the angular acceptance of the SPEC and VTX 
detectors stop in the material of the beam pipe or the first
detector layer.  The multiplicities we report are corrected
for this missing fraction based on the HIJING momentum distributions.

We further studied the contamination of the tracklet distribution 
by feed-down products from weak decays of strange particles. 
Due to the proximity of our detectors to the beamline and 
the good pointing accuracy in the tracklet reconstruction, the contribution
was found to be small ($<4\%$). Again, the multiplicities reported here
are corrected based on the HIJING distributions.

Fig.~\ref{5} shows a direct comparison of the SPEC (left) and VTX (right) 
tracklet $dN/d\eta$ distributions for data (symbols) and MC events 
(solid lines), normalized per event.
Scaling factors of 1.15 for $\sqrt{s} = 56$~AGeV and 
0.98 for $\sqrt{s} = 130$~AGeV were applied to the MC distribution
to match the integrals to  the data.
The shape of the distribution agrees well between simulation and data.
SPEC and VTX distributions are both consistent with the same scaling factor.
Based on the MC studies, the comparison of data and MC tracklet distributions
and the comparison of results from the SPEC and VTX tracklet analysis, we 
conclude that the proportionality factors can be applied to the 
measured tracklet
distributions with an overall systematic uncertainty of less than 8\%.

As the result of this procedure we obtain a primary charged particle density of 
$dN/d\eta |_{|\eta|<1}  = 408 \pm 12 \mbox{(stat)} \pm 30 \mbox{(syst)}$ 
for $\sqrt{s} = 56$~AGeV and $555 \pm 12 \mbox{(stat)} \pm 35 \mbox{(syst)}$ at $\sqrt{s} = 130$~AGeV.
From the simulation of the paddle counter selection we obtain
for the mean number of participating nucleons 
$\langle N_{part} \rangle = 330 \pm 4 \mbox{(stat)} ^{+10}_{-15} 
\mbox{(syst)}$ for
$\sqrt{s} = 56$~AGeV and $343 \pm 4 \mbox{(stat)} 
^{+7}_{-14} \mbox{(syst)}$ for $\sqrt{s} = 130$~AGeV.

Normalizing per participant pair, we deduce
$dN/d\eta |_{|\eta|<1} / 0.5 \langle N_{part} \rangle = 2.47 \pm 0.1 \mbox{(stat)} 
\pm 0.25 \mbox{(syst)}$ and 
$3.24 \pm 0.1 \mbox{(stat)} \pm 0.25 \mbox{(syst)}$, 
respectively.

Finally, taking the strong correlation between the systematic errors at the 
two energies into account, we obtain an increase in the charged particle 
density per participant by a factor of $1.31 \pm 0.04 \mbox{(stat)} \pm 0.05 \mbox{(syst)}$.

\begin{figure}[t]
\centerline{ \epsfig{file=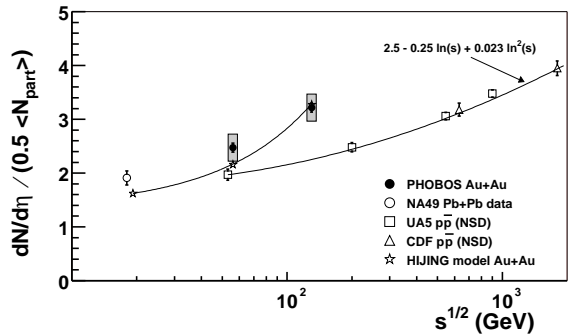,width=8cm} }
\caption{Measured pseudorapidity density normalized per 
participant pair for central Au+Au collisions. Systematic errors
are shown as shaded area. Data are compared with 
$p\overline{p}$ data and Pb+Pb data from the CERN SPS. Also
shown are results of a HIJING simulation (with a line to guide the eye)
and a parametrization of the $p\overline{p}$ data [7].}
\label{6}
\end{figure}

In Fig.~\ref{6} we show the normalized yield per participant obtained
for Au+Au collisions, proton-antiproton ($p\overline{p}$) collisions 
\cite{pp} and 
central Pb+Pb collisions at the CERN SPS \cite{na49}. The 
$dN/d\eta$ value for the Pb+Pb data was obtained
by numerically integrating the momentum distributions shown in \cite{na49}.

Several important features of the data emerge: First,
the central Au+Au collisions show a significantly larger
charged particle density per participant 
than for example  non-single diffractive (NSD) $p\overline{p}$ collisions at comparable 
energies. 
This rules out simple superposition models such as the wounded nucleon 
model \cite{wounded} and is compatible with predictions of models 
like HIJING that include particle production via hard-scattering processes.

Secondly, the observed increase by 31\% from 56 to 130~AGeV in
central Au+Au collisions is significantly steeper than the 
increase shown by a $p\overline{p}$ parametrization (see Fig.~\ref{6})
for the same energy interval \cite{pp}.
Finally, comparing our data to those obtained at the 
CERN SPS for Pb+Pb collisions at $\sqrt{s} = 17.8$~AGeV, 
we find a 70\% higher particle density per participant near $\eta = 0$
at $\sqrt{s} = 130$~AGeV.  
General arguments (c.f. Bjorken's estimate \cite{Bjorken}) suggest 
that this increase should correspond to a similar increase in the maximal energy density 
achieved in the collision.

Acknowledgements: 
We acknowledge the generous support of the entire RHIC project personnel, C-A
and Chemistry Departments at BNL. We thank Fermilab and CERN for help in
silicon detector assembly. We thank the MIT School of Science and LNS for
financial
support. This work was partially supported by US DoE grants DE-AC02-98CH10886,
DE-FG02-93ER40802, DE-FC02-94ER40818, DE-FG02-94ER40865, DE-FG02-99ER41099, W-31-109-ENG-38.
NSF grants 9603486, 9722606 and 0072204. The Polish groups were partially supported by KBN grant 2 P03B
04916. The NCU group was partially supported by NSC of Taiwan under 
contract NSC 89-2112-M-008-024.

\end{document}